\documentclass[aps,twocolumn,amsmath,amssymb,prb,superscriptaddress]{revtex4}
\usepackage[dvipdfmx]{graphicx}
\usepackage[dvipdfmx]{color}
\usepackage{amsmath}
\usepackage{amssymb}
\usepackage{dcolumn}
\usepackage{bm}
\usepackage{latexsym} 
\usepackage{setspace}
\usepackage{graphicx}
\usepackage{color}
\usepackage[dvipdfmx,colorlinks=true,linkcolor=blue,citecolor=blue]{hyperref}%2023/5/27 added; linkcolor sets the color of ``\ref'', while citecolor sets the color of ``\cite''.
\begin{document}
\title{Cryogenic spin Peltier effect detected by a RuO$_2$-AlO$_x$ on-chip microthermometer} 
\author{Takashi Kikkawa}
\email{t.kikkawa@ap.t.u-tokyo.ac.jp}
\affiliation{Department of Applied Physics, The University of Tokyo, Tokyo 113-8656, Japan}
\author{Haruka Kiguchi}
\affiliation{Department of Physics, Ochanomizu University, Tokyo 112-8610, Japan}
\affiliation{Department of Applied Physics, The University of Tokyo, Tokyo 113-8656, Japan}
\author{Alexey A. Kaverzin}
\affiliation{Department of Applied Physics, The University of Tokyo, Tokyo 113-8656, Japan}
\affiliation{Institute for AI and Beyond, The University of Tokyo, Tokyo 113-8656, Japan}
\author{Ryo Takahashi}
\affiliation{Department of Physics, Ochanomizu University, Tokyo 112-8610, Japan}
\author{Eiji Saitoh}
\affiliation{Department of Applied Physics, The University of Tokyo, Tokyo 113-8656, Japan}
\affiliation{Institute for AI and Beyond, The University of Tokyo, Tokyo 113-8656, Japan}
\affiliation{WPI Advanced Institute for Materials Research, Tohoku University, Sendai 980-8577, Japan}
%\affiliation{Center for Spintronics Research Network, Tohoku University, Sendai 980-8577, Japan}
%\affiliation{Advanced Science Research Center, Japan Atomic Energy Agency, Tokai 319-1195, Japan}
%
\date{\today}
\begin{abstract} 
We report electric detection of the spin Peltier effect (SPE) in a bilayer consisting of a Pt film and a Y$_{3}$Fe$_5$O$_{12}$ (YIG) single crystal at the cryogenic temperature $T$ as low as 2 K based on a RuO$_2$$-$AlO$_x$ on-chip thermometer film.  
By means of a reactive co-sputtering technique, we successfully fabricated RuO$_2$$-$AlO$_x$ films having a large temperature coefficient of resistance (TCR) of $\sim 100\% ~\textrm{K}^{-1}$ at around $2~\textrm{K}$. 
By using the RuO$_2$$-$AlO$_x$ film as an on-chip temperature sensor for a Pt/YIG device, we observe a SPE-induced temperature change on the order of sub-$\mu \textrm{K}$, the sign of which is reversed with respect to the external magnetic field $B$ direction.  
We found that the SPE signal gradually decreases and converges to zero by increasing $B$ up to $10~\textrm{T}$.
The result is attributed to the suppression of magnon excitations due to the Zeeman-gap opening in the magnon dispersion of YIG, whose energy much exceeds the thermal energy at 2 K.
\end{abstract} 
\maketitle
% 
%
%------------main-text----------------------------------
%
%%%%%%%%%%%%%%%%%%%%%%%%%%%%%%%%%%%%%%%%%%%%%%%%%%%%
\section{INTRODUCTION}
%\section{I.~~INTRODUCTION} 
One of the important features in spintronics is that various phenomena have been found at room temperature in simple stacked structures, leading to their practical device applications \cite{Ohno2016ProcIEEE,Chumak2022IEEETransMagn,Yang2022Nature,Maekawa2023JAP}.  
Meanwhile, exploring the spintronic phenomena at low temperatures often resulted in a discovery of new functional properties with both fundamental and practical prospects \cite{Fan2014NatMat,Linder2015NatPhys,Umeda2018APL,Yao2018PRB,Shiomi2019NatPhys,Jeon2020ACSNano,Tschirhart2023NatPhys}.  
A typical example is the spin Seebeck effect (SSE), which refers to the generation of a spin current as a result of a temperature gradient in a magnetic material, and has been observed at room temperature in a variety of magnetic materials, including garnet- and spinel-ferrites with high magnetic ordering temperatures\cite{Uchida2016ProcIEEE,Rezende_text,Kikkawa2023ARCMP}. 
When SSEs are measured at low temperatures in certain systems however, intriguing physics comes to the surface. 
Major findings include the signal anomalies induced by hybridized magnon-phonon excitations\cite{Kikkawa2023ARCMP,Kikkawa2016PRL,Cornelissen2017PRB,Oyanagi2020AIPAdv,Li2020PRL_Cr2O3,Kikkawa2022PRMater}, unconventional sign reversal due to competing magnon modes having opposite spin polarizations\cite{Geprags2016NatCommun}, observation of a spin-superfluid-mediated nonlocal SSE signal\cite{Yuan2018SciAdv}, and SSEs driven by paramagnetic spins\cite{Wu2015PRL,Oyanagi2023PRB}  and exotic elementary excitations in quantum spin systems\cite{Hirobe2017NatPhys,Chen2021NatCommun,Xing2022APL}.  
Furthermore, recently, a nuclear SSE has been observed in an antiferromagnet having strong hyperfine coupling\cite{Kikkawa2021NatCommun,Kikkawa2023ARCMP}. 
The signal increases down to ultralow temperatures on the order of $100~\textrm{mK}$, which is distinct from conventional thermoelectric effects in electronic (spin) systems\cite{Kikkawa2023ARCMP,Kikkawa2021NatCommun}, and may offer an opportunity for exploring thermoelectric science and technologies at ultralow temperatures, an important environment in quantum information science. \par
%
%remains unexplored
In contrast to the intense research on SSEs, the spin Peltier effect \cite{Flipse2014PRL,Daimon2016NatCommun,Daimon2017PRB,Itoh2017PRB,Yagmur2018JPhysD,Sola2019SciRep,Yahiro2020PRB,Yamazaki2020PRB,Daimon2020APEX,Uchida2021Review,Uchida2021JPSJ,Takahagi2023APL}, the reciprocal of the SSE, remains to be explored at low temperatures below 100 K because of its experimental difficulty.
The SPE modulates the temperature of a junction consisting of a metallic film and a magnet in response to a spin current\cite{Daimon2016NatCommun}, and has been detected usually by means of lock-in thermography (LIT) \cite{Daimon2016NatCommun,Daimon2017PRB,Yagmur2018JPhysD,Daimon2020APEX} and thermocouples \cite{Flipse2014PRL,Itoh2017PRB,Yahiro2020PRB}.  
The LIT measures the infrared intensity emitted from the sample surface based on a combination of the lock-in with temperature imaging technique, whose intensity is in proportion to the fourth power of the absolute temperature $T$ (the Stefan--Boltzmann law\cite{Yagmur2018JPhysD,Uchida2021JPSJ}).
This results in a typical resolution of 0.1 mK at room temperature \cite{Uchida2021JPSJ}, which is sufficient to measure a SPE in a prototypical Pt/Y$_{3}$Fe$_5$O$_{12}$ (YIG) system at higher temperatures ($\sim$ room temperature and above) \cite{Daimon2016NatCommun,Daimon2017PRB}. 
However, the LIT may not be applicable for detecting the low-temperature SPE, because its sensitivity is dramatically reduced with decreasing temperature \cite{Yagmur2018JPhysD,Uchida2021JPSJ}. 
Furthermore, a thermocouple micro-sensor with a high resolution of $\sim 5~\mu\textrm{K}$ was used to measure a SPE down to 100 K  in Ref. \onlinecite{Yahiro2020PRB}. However, it was found to be difficult to conduct the measurements below 100 K as the sensitivity of the thermocouple decreases with decreasing $T$.
It is therefore important to establish an alternative experimental method for detecting cryogenic SPEs\cite{Maekawa2023JAP}. 
An ultimate goal in this direction would be to find cryogenic SPEs driven by nuclear and quantum spins that can be activated even at ultralow temperatures, toward future possible cooling- and heat-pump technologies in such an environment. \par 
In this study, we have explored the SPE at a cryogenic temperature below the liquid-$^4$He temperature in a prototypical Pt/YIG system.  %based on an on-chip micro-thermometer. 
There are three crucial requirements for practical realization of such measurement that are (1) the high temperature-resolution of $\sim$ sub-$\mu\textrm{K}$-order or better at low temperatures, (2) ability to detect a temperature change of a metallic (Pt) thin film (which implies for contact-mode measurements sufficient thermal coupling and low heat capacity), and (3) reliability under a high magnetic-field environment.  
To realize the thermometry that meets these requirements, we adopted a RuO$_2$-based micro-thermometer \cite{Pobell-textbook,LakeShore,Bosch1986Cryogenics,Li1986Cryogenics,Batko1995Cryogenics,Neppert1996Cryogenics,Affronte1997JLowTempPhys,Chen2003,Chen2009,Nelson2015RevSciInstrum}
 (RuO$_2$$-$AlO$_x$ composite film in our case).
In general, RuO$_2$-based resistors show a high temperature-sensitivity due to their large negative temperature coefficient of resistance. Besides, they show reasonably small magnetoresistance and can be made in a thin-film form. Owing to these advantages, in fact, RuO$_2$-based chip resistors have widely been used as temperature sensors at cryogenic temperatures \cite{Pobell-textbook,LakeShore}. 
We have fabricated RuO$_2$$-$AlO$_x$ films by means of a co-sputtering technique and found the optimal fabrication condition by characterizing their electric transport properties. 
By using a RuO$_2$$-$AlO$_x$ film as an on-chip temperature sensor for a Pt-film/YIG-slab system, we successfully measured a SPE-induced temperature change on the order of sub-$\mu \textrm{K}$ at $T = 2~\textrm{K}$.  
Our results provide an important step toward a complete physical picture of the SPE and establishment of cryogenic spin(calori)tronics \cite{Uchida2021Review}. \par  
%
%%%%%%%%%%%%%%%%%%%%%%%%%%%%%%%%%%%%%%%%%%%%%%%%%%%%
\section{EXPERIMENTAL PROCEDURE} \label{sec:procedure}
%\section{II.~~EXPERIMENTAL PROCEDURE} \label{sec:procedure}
%%%%%%%%%%%%%%%%%%%%%%%%%%%%%%%%%%%%%%%%%%%%%%%%%%%%
%
%%%%%%%%%%%%%%%%%%%%%%%%%%%%%%%%%%%%%%%%%%%%%%%%%%%%
\subsection{Fabrication of RuO$_2$$-$AlO$_x$ films} \label{sec:fabrication}
%%%%%%%
%
%
%
%
We have fabricated RuO$_2$$-$AlO$_x$ composite films as a micro-thermometer by means of d.c. co-sputtering technique from RuO$_2$ (99.9\%, 2-inch diameter) and Al (99.999\%, 2-inch diameter) targets under Ar and O$_2$ atmosphere.  
%RuO2 (5 mm in thick): Toshima Manufacturing Co., Ltd.
%Al (3 mm in thick): Kojundo Chemical Lab. Co., Ltd.
%Here, 5-nm-thick Pt films were deposited at a d.c.-power of 20 W and an Ar pressure of 0.10 Pa at the rate of 0.0216 nm sec$^{-1}$.
%Pt depo. rate = 5nm/231sec, pre-sputtering time = 60 sec, Ar 15sccm
%Al pre-sputtering condition before RuOx-AlOx co-sputtering: d.c. 30W, pre-sputtering time = 600 sec, Ar 15sccm. 
To obtain the most suitable thermometer film for the SPE at low temperatures, a series of co-sputtered RuO$_2$$-$AlO$_x$ films on thermally-oxidized Si substrates  was first prepared at several d.c. power values for the RuO$_2$ target ($P_{{\rm RuO}_2}=25$, $26$, $27$, $28$, and $30~\textrm{W}$) and the fixed d.c. power for the Al target ($P_{{\rm AlO}_x}=25~\textrm{W}$) under a sputtering gas of Ar + 7.83 vol.\% O$_2$ at a pressure of 0.13 Pa at room temperature. 
%0.13 Pa -> This value is based on the deposition on Feb. 18th, 2023, see the logbook. 
%Ar: 15.3 sccm, O2: 1.3 sccm -> 1.3/(15.3 + 1.3) = 0.0783 = 7.83%
Here, the values of Ar$-$O$_2$ gas amount and the d.c. power of $P_{{\rm AlO}_x}=25~\textrm{W}$ were chosen such that highly-insulating AlO$_x$ films are obtained with a reasonable deposition rate ($\sim 1~\textrm{nm/min}$) when AlO$_x$ is sputtered solely from the Al target. 
We note that, if the O$_2$ gas amount exceeds an onset value, the d.c. sputtering rate suddenly decreases due to the surface oxidization of the Al target \cite{Maniv1980AlOx,Wallin2008AlOx}, whereas if the O$_2$ gas amount is insufficient, the resultant AlO$_x$ film may show finite electrical conduction. 
We found that the introduction of O$_2$ by itself does not play an important role in the temperature variation of the resistance for pure RuO$_2$ films (for details, see APPENDIX A). To keep the sputtering conditions and resultant films' quality as consistent as possible through repeated deposition cycles, we introduced common pre-sputtering processes just before actual depositions. To remove a possible oxidized top layer of the Al target, it was pre-sputtered at a relatively high power of $P_{{\rm AlO}_x}=30~\textrm{W}$ for 600 s without introducing O$_2$ gas, and then the RuO$_2$ and Al targets were pre-sputtered for 60 s under the actual deposition conditions (i.e., Ar + 7.83 vol.\% O$_2$) \cite{Comment_maintenance}.
%which were performed just before actual depositions.
For electric transport measurements of the RuO$_2$$-$AlO$_x$ films, they were patterned into a Hall-bar shape having the length, width, and thickness of $1.0~\textrm{mm}$, $0.5~\textrm{mm}$, and $\sim 100~\textrm{nm}$, respectively, by co-sputtering RuO$_2$$-$AlO$_x$ through a metal mask. 
The RuO$_2$ content in the RuO$_2$$-$AlO$_x$ films under the different RuO$_2$ sputtering power $P_{{\rm RuO}_2}$  was evaluated through scanning electron microscopy with energy dispersive X-ray analysis (SEM-EDX) and the surface roughness of the films was characterized through atomic force microscopy (AFM).  
%Ar 15.3sccm, O2 1.3 sccm
\par
\subsection{Fabrication of SPE device} \label{sec:Fabrication_SPE}
\begin{figure}[htb]
\begin{center}
\includegraphics[width=8.5cm]{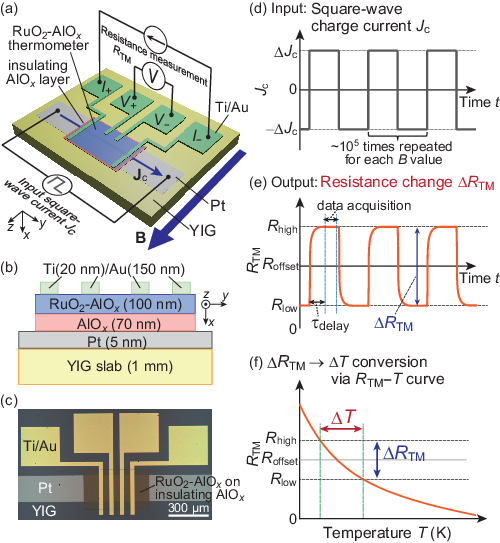}
\end{center}
\caption{(a) A schematic illustration of the SPE device consisting of a Pt-film/YIG-slab bilayer, on top of which a RuO$_2$$-$AlO$_x$ thermometer (TM) film is attached for the detection of the SPE-induced temperature change $\Delta T$ in the Pt film. 
Besides, in this device, Au/Ti electrodes are formed on the RuO$_2$$-$AlO$_x$ film for the 4 terminal resistance measurements and an AlO$_x$ film is inserted between the RuO$_2$$-$AlO$_x$ and Pt films for the electrical insulation between them. 
%To perform 4 terminal resistance measurements, Au/Ti electrodes are formed on the RuO$_2$$-$AlO$_x$ film for the 4 terminal resistance measurements and to electrically isolate the RuO$_2$$-$AlO$_x$ and Pt films an AlO$_x$ film is inserted between them. 
(b) A schematic side-view image of the SPE device, where the numbers in parentheses represent the thickness.
(c) An optical microscope image of a typical SPE device. 
%The white scale bar corresponds to $300~\mu\textrm{m}$.
(d) Input signal: A square-wave charge current $J_{\rm c}$ with amplitude $\Delta J_{\rm c}$ applied to the Pt film. 
(e) Output signal: A resistance $R_{\rm TM}$ in the RuO$_2$$-$AlO$_x$ film that responds to the change in the $J_{\rm c}$ polarity, $\Delta R_{\rm TM} (\equiv R_{\rm high}-R_{\rm low})$ originating from the SPE-induced $\Delta T$ of the Pt film ($\propto \Delta J_{\rm c}$) \cite{Itoh2017PRB,Yahiro2020PRB}. Here, the Joule-heating-induced temperature change ($\propto \Delta J_{\rm c}^2$) is constant in time, and does not overlap with $\Delta R_{\rm TM}$. 
(f) A schematic illustration of the temperature $T$ dependence of $R_{\rm TM}$, from which the $\Delta R_{\rm TM}$ value can be converted to the temperature change $\Delta T$. 
}
\label{fig:1}
\end{figure}

To investigate the SPE below the liquid-$^4$He temperature, we have prepared devices consisting of a Pt-film/YIG-slab bilayer, where a 100-nm-thick RuO$_2$$-$AlO$_x$ film with Au/Ti electrodes is attached on the top surface of the Pt film to detect its SPE-induced temperature change $\Delta T$ [see the schematic illustrations and the optical microscope image of a typical SPE device shown in Figs. \ref{fig:1}(a) -- \ref{fig:1}(c)].  
Three photolithography steps were employed to make the SPE devices, where all the film depositions were performed at room temperature. 
First, a 5-nm-thick Pt wire with the width of $200~\mu \textrm{m}$ was formed on the (111) surface of a single-crystalline YIG slab with the size of $5\times5\times1~\textrm{mm}^3$ by d.c. magnetron sputtering in a $0.1~\textrm{Pa}$ Ar atmosphere under the d.c. power of 20 W. %, resulting in a deposition rate of 0.022 nm sec$^{-1}$. 
In the next photolithography step, a 70-nm-thick insulating AlO$_x$ layer was formed at the area of $230 \times 350~\mu \textrm{m}^2$ [$300 \times 500~\mu \textrm{m}^2$ for the device shown in Fig. \ref{fig:1}(c)] on top of the Pt/YIG layer to electrically isolate the RuO$_2$$-$AlO$_x$ film from the Pt layer. 
Here, the AlO$_x$ deposition was done by r.f. magnetron sputtering from an Al$_2$O$_3$ target (99.99\%, 2-inch diameter) under the r.f. power of 150 W  and a sputtering gas of Ar + 1.0 vol.\% O$_2$ \cite{Voigt2004AlOx} at a pressure of 0.6 Pa. We later confirmed that the AlO$_x$ film shows a high electric resistance on the order of $1-10~\textrm{G}\Omega$ along the out-of-plane direction at room temperature. 
%resulting in a deposition rate of $\sim$ 0.038888 nm sec$^{-1}$.  
%Ar: 3sccm, O2: 0.03 sccm (digital display of K-Science sputtering machine)
%(YIG commercially available from SurfaceNet GmbH) 
%Pt depo rate = 5 nm/(231 sec) = 0.216 nm/sec
%Al2O3 depo rate = 70 nm/(30 min) = 2.33333 nm/min = 0.0388888888 nm/sec, 4.6E-3 Torr = 0.613 Pa
Subsequently, a 100-nm-thick RuO$_2$$-$AlO$_x$ thermometer film was deposited on top of the AlO$_x$ layer at the area of $230 \times 350~\mu \textrm{m}^2$ [$300 \times 500~\mu \textrm{m}^2$ for the device shown in Fig. \ref{fig:1}(c)] through the co-sputtering under the d.c. sputtering power of $P_{{\rm RuO}_2} = 28~\textrm{W}$ and $P_{{\rm AlO}_x} = 25~\textrm{W}$. 
Here, the dimensions and sputtering power for the RuO$_2$$-$AlO$_x$ film were chosen such that the resistance $R$ of the resulting film is several tens of k$\Omega$ at 2 K and its sensitivity monotonically increases with decreasing $T$ down to $2~\textrm{K}$\cite{Comment_different-T} [as shown in Figs. \ref{fig:3} and \ref{fig:4}(d) and discussed in Sec. \ref{sec:Electric_conduction}].  
%
%%%%%%%%%%%%%%%%%%%%%%%%%%%%%%%%%%%%%%%%%%%%%%%%%%%%%%%%%%%%%%%%%%
%%%%%%%%%%%%%%%%%%%%%%%%%%%%%%%%%%%%%%%%%%%%%%%%%%%%%%%%%%%%%%%%%%  
%
We then proceeded with the final photolithography step for Au(150 nm)/Ti(20 nm) electrodes, where the numbers in parentheses represent the thicknesses of the deposited films. Each Au/Ti electrode wire on the RuO$_2$$-$AlO$_x$ film has the $30$-$\mu \textrm{m}$ width and is placed at $50$-$\mu \textrm{m}$ intervals. 
To reduce the contact resistance between the RuO$_2$$-$AlO$_x$ and Ti films, Ar-ion milling was performed directly before depositing the Au/Ti film.  
Both the Ti and Au layers were formed by r.f. magnetron sputtering in succession without breaking vacuum.  
The first lithography process for the Pt layer was done using a single-layer photoresist (AZ5214E) followed by a lift-off process, whereas the second and third processes for the AlO$_x$/RuO$_2$$-$AlO$_x$ and Au/Ti layers were done using a double-layered photoresist (LOR-3A and AZ5214E)  to provide an undercut structure for a better success rate of the lift off process.   
\par
%generates a spin current ${\bf J}_{\rm s}=J_{\rm s}\mathbf{\hat{x}}$ 
%{\it Principle of SPE and its measurement.}
%
%
%
%
\subsection{SPE and SSE measurements} \label{sec:Method_SPE-SSE}
Figure \ref{fig:1}(a) shows a schematic illustration of the SPE device and the experimental setup in the present study. 
The SPE appears as a result of the interfacial spin and energy transfer between magnons in YIG and electron spins in Pt \cite{Flipse2014PRL,Daimon2016NatCommun,Daimon2017PRB,Itoh2017PRB}.   
Suppose that the magnetization ${\bf M}$ of the YIG layer is oriented along the $+ \mathbf{\hat{z}}$ direction by the external magnetic field ${\bf B} \;||+\mathbf{\hat{z}}$, as shown in Fig. \ref{fig:1}(a).
With the application of a charge current ${\bf J}_{\rm c} =J_{\rm c}\mathbf{\hat{y}}$ to the Pt film, the spin Hall effect (SHE)\cite{SHE_Hoffman,SHE_Sinova} induces a nonequilibrium spin, or magnetic moment, accumulation at the Pt/YIG interface \cite{Flipse2014PRL,Daimon2016NatCommun,Daimon2017PRB,Itoh2017PRB,Yahiro2020PRB}.
%Here, based on the nature of SHE in Pt and the antiparallel orientation between the electron spin and corresponding magnetic moment, 
For ${\bf J}_{\rm c} \;||+\mathbf{\hat{y}}$ (${\bf J}_{\rm c} \;||-\mathbf{\hat{y}}$), the accumulated magnetic moment $\delta {\bf m}_{\rm s}$ at the interfacial Pt orients along the $- \mathbf{\hat{z}}$ ($+ \mathbf{\hat{z}}$) direction \cite{Schreier2014JPhysD,Daimon2017PRB}, which is antiparallel (parallel) to the ${\bf M}$ direction in Fig. \ref{fig:1}(a).  
Through the interfacial spin-flip scattering, $\delta {\bf m}_{\rm s}$ creates or annihilates a magnon in YIG; the number of magnons in YIG increases (decreases) when $\delta {\bf m}_{\rm s} \;||-{\bf M}$ ($\delta {\bf m}_{\rm s} \;||\;{\bf M}$)\cite{Uchida2021Review,Maekawa2023JAP}. 
Because of energy conservation, this process is accompanied by a heat flow  ${\bf J}_{\rm q}$ between the electron in Pt and the magnon in YIG\cite{Uchida2021Review,Maekawa2023JAP}. 
The temperature of Pt (YIG) thus decreases (increases) when $\delta {\bf m}_{\rm s} \;||-{\bf M}$ under ${\bf J}_{\rm c} \;||+\mathbf{\hat{y}}$ and ${\bf B} \;||+\mathbf{\hat{z}}$ [Fig. \ref{fig:1}(a)], whereas the temperature of Pt (YIG)  increases (decreases) when $\delta {\bf m}_{\rm s} \;||\;{\bf M}$ by reversing either ${\bf J}_{\rm c}$ or ${\bf B}$ in Fig. \ref{fig:1}(a)\cite{Daimon2016NatCommun,Daimon2017PRB,Itoh2017PRB,Yahiro2020PRB}. The SPE-induced temperature change $\Delta T$ satisfies the following relationship \cite{Daimon2016NatCommun,Daimon2017PRB,Itoh2017PRB}
\begin{equation}
\Delta T \; \propto \; \delta {\bf m}_{\rm s} \cdot {\bf M} \; \propto \;  ({\bf J}_{\rm c} \times {\bf M}) \cdot \mathbf{\hat{x}}.
\label{eq:SPE}
\end{equation}
\par
For the electric SPE detection based on the on-chip thermometer (TM), we utilized the highly-accurate resistance measurement scheme called the Delta mode, a combination of low-noise current source and nanovoltmeter (Keithley  Model 6221 and 2182A \cite{Itoh2017PRB,Yahiro2020PRB}).
%To this end, 
We applied a square-wave charge current $J_{\rm c}$ with amplitude $\Delta J_{\rm c}$ to the Pt film [Figs. \ref{fig:1}(a) and \ref{fig:1}(d)] and measured the 4 terminal RuO$_2$$-$AlO$_x$ resistance $R_{\rm TM}$ that responds to the change in the $J_{\rm c}$ polarity, $\Delta R_{\rm TM} \equiv R_{\rm high}-R_{\rm low}$, where $R_{\rm high}$ ($R_{\rm low}$) represents the $R_{\rm TM}$ value for $J_{\rm c} = + \Delta J_{\rm c}$ $(- \Delta J_{\rm c})$ and was measured under the sensing current of $100~\textrm{nA}$ applied to the RuO$_2$$-$AlO$_x$ film [see Figs. \ref{fig:1}(a) and \ref{fig:1}(e)] \cite{Itoh2017PRB}. 
Here, the $\Delta R_{\rm TM}$ value is free from the Joule-heating-induced resistance change ($\propto \Delta J_{\rm c}^2$) that is independent of time, which thereby only contributes to the offset resistance $R_{\rm offset}$ of the RuO$_2$$-$AlO$_x$ film shown in Fig. \ref{fig:1}(e) \cite{Comment_Joule-heating}. 
During the SPE measurement, the magnetic field ${\bf B}$ (with magnitude $B$) was applied in the film plane and perpendicular to the Pt wire, i.e., ${\bf B} \;||\; \mathbf{\hat{z}}$ in Fig. \ref{fig:1}(a), except for the control experiment shown in Fig. \ref{fig:4}(b), where ${\bf B} \;||\; \mathbf{\hat{x}}$.
The resistance $R_{\rm high,low}$ was recorded after the time delay $\tau_{\rm delay}$ of $50~\textrm{ms}$ [except for the $\tau_{\rm delay}$ dependence shown in Fig. \ref{fig:4}(e)] during the data acquisition time $\tau_{\rm sens}$ of $20~\textrm{ms}$ \cite{Comment_frequency}, and then was accumulated by repeating the process of the $J_{\rm c}$-polarity change $\sim 10^5$ times for each $B$ point [see Fig. \ref{fig:1}(d)] to improve the signal-to-noise ratio. 
$\Delta R_{\rm TM}$ can be converted into the corresponding temperature change $\Delta T ~(= \Delta R_{\rm TM}/S)$ by using the sensitivity $S \equiv |dR_{\rm TM}/dT|$ of the RuO$_2$$-$AlO$_x$ film [see Figs. \ref{fig:1}(f) and \ref{fig:4}(d)]. \par 
%All the $\Delta R_{\rm TM}$ and $\Delta T$ data were anti-symmetrized with respect to the magnetic field $B$. \par 
%
%B is parallel to the $[1\overline{1}0]$ direction of the YIG slab.
%
To compare the $B$ dependence of the SPE signal with that of the SSE, we also measured the SSE at $T=2~\textrm{K}$ using the same device, for which all SPE results presented in this paper were obtained, but in a different experimental run from the SPE measurement. 
Here, the SSE measurement was done by means of a lock-in detection technique \cite{Wu2015PRL,Kikkawa2022PRMater,Kikkawa2021NatCommun} and the RuO$_2$$-$AlO$_x$ layer was used as a resistive heater; an a.c. charge current $I_{\rm c}=\sqrt{2}I_{\rm rms} {\rm sin}(\omega t)$ with the amplitude of $I_{\rm rms} = 5.48 ~\mu{\rm A}$ and the frequency of $\omega/2\pi = 13.423~\textrm{Hz}$ was applied to the RuO$_2$-AlO$_x$  film, and the second harmonic voltage in the Pt layer induced by a spin current (driven by a heat current due to the Joule heating of the RuO$_2$-AlO$_x$  film $P_{\rm heater} = R_{\rm TM} I_{\rm rms}^2$) was detected. %
During the SSE measurement, the external field ${\bf B}$ was applied in the film plane and perpendicular to the Pt wire, i.e., ${\bf B} \;||\; \mathbf{\hat{z}}$ in Fig. \ref{fig:1}(a). \par
%
%%%%%%%%%%%%%%%%%%%%%%%%%%%%%%%%%%%%%%%%%%%%%%%%%%%%
\section{RESULTS AND DISCUSSION}
%\section{III.~~RESULTS AND DISCUSSION}
%%%%%%%%%%%%%%%%%%%%%%%%%%%%%%%%%%%%%%%%%%%%%%%%%%%%
%
%
\subsection{Electrical conduction in RuO$_2$$-$AlO$_x$ films} \label{sec:Electric_conduction} 
\begin{figure}[htb]
\begin{center}
\includegraphics[width=8.5cm]{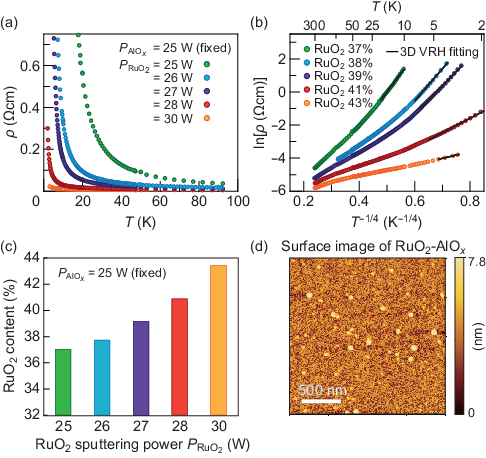}
\end{center}
\caption{
(a) $T$ dependence of the resistivity $\rho$  for the RuO$_2$$-$AlO$_x$ films fabricated on thermally-oxidized Si substrates under the several d.c. sputtering power for the RuO$_2$ target ($P_{{\rm RuO}_2}$) and the fixed d.c. power for the Al target ($P_{{\rm AlO}_x}$).
(b) ${\rm ln} \rho$ versus $T^{-1/4}$ for the RuO$_2$$-$AlO$_x$ films. The black solid lines are obtained by fitting Eq. (\ref{eq:3D-Mott-VRH}) (the 3D Mott VRH model) to the experimental data. 
(c) Relationship between the RuO$_2$ content in the RuO$_2$$-$AlO$_x$ films and the RuO$_2$ sputtering power $P_{{\rm RuO}_2}$ determined by SEM-EDX. 
% from which the figure legends describing the RuO$_2$ content in Fig. \ref{fig:2}(b) and Fig. \ref{fig:3} are obtained.
% 
Using this correspondence, the figure legends in (b) and also Fig. \ref{fig:3} are described in terms of the RuO$_2$ content. 
% Using this correspondence, the figure legends based on $P_{{\rm RuO}_2}$ values in (a) can be replaced with those based on the RuO$_2$ content. This representation is adopted in Fig. \ref{fig:2}(b) and Fig. \ref{fig:3}. 
(d) A typical AFM image of the RuO$_2$$-$AlO$_x$ film grown under $P_{{\rm RuO}_2} = 28~\textrm{W}$ and $P_{{\rm AlO}_x} = 25~\textrm{W}$ (the RuO$_2$ content of 41\%), where the root-mean-squared surface roughness is $R_{\rm rms} = 1.2~\textrm{nm}$. The white scale bar represents 500 nm.
}
\label{fig:2}
\end{figure}

We first characterize the electrical conduction of the RuO$_2$$-$AlO$_x$ films on thermally-oxidized Si substrates. 
Figure \ref{fig:2}(a) shows the $T$ dependence of the resistivity $\rho$ for the films grown under the several (fixed) sputtering power values for the RuO$_2$ (Al) target $P_{{\rm RuO}_2}$ ($P_{{\rm AlO}_x}$). 
For all the films, $\rho$ increases with decreasing $T$ in the entire temperature range, showing a negative temperature coefficient of resistance (TCR). 
Both $\rho$ and its slope $|d\rho/dT|$ increase significantly at low temperatures and monotonically by decreasing the RuO$_2$ sputtering power $P_{{\rm RuO}_2}$.  
The overall $\rho$--$T$ curve shifts toward the upper right by decreasing $P_{{\rm RuO}_2}$. 
The result shows that the $\rho$ versus $T$ characteristics of the RuO$_2$$-$AlO$_x$ films can be controlled simply by changing the sputtering power $P_{{\rm RuO}_2}$. 
SEM-EDX analysis reveals that the RuO$_2$/AlO$_x$ ratio decreases by decreasing $P_{{\rm RuO}_2}$ [Fig. \ref{fig:2}(c)], which leads to the $\rho$ increase in the electric transport. 
We also characterized the RuO$_2$$-$AlO$_x$ films by means of AFM and found that a typical root-mean-squared surface roughness is $R_{\rm rms} \sim 1~\textrm{nm}$, much smaller than their thickness $\sim 100~\textrm{nm}$ [see the AFM image of the RuO$_2$$-$AlO$_x$ films grown under $P_{{\rm RuO}_2} = 28~\textrm{W}$ and $P_{{\rm AlO}_x} = 25~\textrm{W}$ (the RuO$_2$ content of 41\%) shown in Fig. \ref{fig:2}(d)]. 
\par
% 
%%%%%%%%%%%%%%%%%%%%%%%%%%%%%%%%%%%%%%%%%%%%%%%%%%%%%%%%%%%%%%%%%%%
%%%%%%%%%%%%%%%%%%%%%%%%%%%%%%%%%%%%%%%%%%%%%%%%%%%%%%%%%%%%%%%%%% 
% where the surface roughness is $R_a = 0.96~\textrm{nm}$.
%
%%%%%%%%%%%%%%%%%%%%%%%%%%%%%%%%%%%%%%%%%%%%%%%%%%%%%%%%%%%%%%%%%%%
%%%%%%%%%%%%%%%%%%%%%%%%%%%%%%%%%%%%%%%%%%%%%%%%%%%%%%%%%%%%%%%%%% 
%
%
%
The electrical conduction at sufficiently low temperatures for RuO$_2$-based thermometers has often been analyzed by the variable-range hopping (VRH) model for three dimensional (3D) systems proposed by Mott \cite{Pobell-textbook,Mott1969PhilosMag,Ambegaokar1971PRB,Bosch1986Cryogenics,Li1986Cryogenics,Batko1995Cryogenics,Neppert1996Cryogenics,Affronte1997JLowTempPhys}, 
\begin{equation}
\rho = \rho_0 {\rm exp}\left(\frac{T_0}{T}\right)^{1/4}, 
\label{eq:3D-Mott-VRH}
\end{equation}
where $\rho_0$ is the resistivity coefficient and $T_0$ is the characteristic temperature related to the electron localization length $a$.  
To discuss our result  in light of the VRH, we plot ${\rm ln} \rho$ versus $T^{-1/4}$ for the RuO$_2$$-$AlO$_x$ films in Fig. \ref{fig:2}(b). 
We found that ${\rm ln} \rho$ scales linearly with $T^{-1/4}$ at low-$T$ ranges, and the ${\rm ln} \rho$--$T^{-1/4}$ data is well fitted by Eq. (\ref{eq:3D-Mott-VRH}) [see the black solid lines in Fig. \ref{fig:2}(b)], suggesting that the low-$T$ electrical conduction is indeed governed by the VRH.    
From the fitting, the $T_0$ values are obtained as 
$2.58\times10^{5}$, $1.41\times10^{5}$, $8.95\times10^{4}$, $5.73\times10^{3}$, and $2.60\times10^{2}~\textrm{K}$
for the RuO$_2$$-$AlO$_x$ films grown under $P_{{\rm RuO}_2}=25$, $26$, $27$, $28$, and $30~\textrm{W}$, respectively. 
We note that, at all the $T$ ranges adopted for the VRH fitting, the average hopping distance ($R_{\rm hop}$) is larger than the electron localization length ($a$) that is the requirement for the VRH model to be valid: $R_{\rm hop}/a = (3/8)(T_0/T)^{1/4}>1$ \cite{Rosenbaum1991PRB,Lafuerza2013PRB,Lu2014APL,Chen2021JPhysD,Li2022JMaterSci}. Besides, the Mott hopping energy $E_{\rm hop} =  (1/4)k_{\rm B}T(T_0/T)^{1/4}$ ($k_{\rm B}$: the Boltzmann constant) obtained for the present films is larger than (or comparable to) the thermal energy $k_{\rm B}T$, allowing for the electron hopping \cite{Rosenbaum1991PRB,Lafuerza2013PRB,Lu2014APL,Chen2021JPhysD,Li2022JMaterSci}. 
The above argument further confirms the validity of the 3D Mott VRH model to describe the conduction mechanism in the RuO$_2$$-$AlO$_x$ films. \par 
%   259.5252594 K (0.0026 E5)for 25w-30w
%  5725.289373 K (0.0573 E5) for 25w-28w
% 89512.38766 K (0.8951 E5) for 25w-27w
%140627.4188 K (1.4063 E5) for 25w-26w
%258208.0523 K (2.5821 E5) for 25w-25w
% 
%
%%%%%%%%%%%%%%%%%%%%%%%%%%%%%%%%%%%%%%%%%%%%%%%%%%%%%%%%%%%%%%%%%%%
%%%%%%%%%%%%%%%%%%%%%%%%%%%%%%%%%%%%%%%%%%%%%%%%%%%%%%%%%%%%%%%%%% 
\begin{figure}[htb]
\begin{center}
\includegraphics[width=8.5cm]{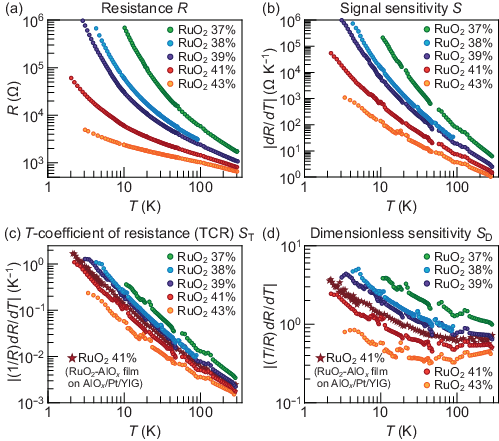}
\end{center}
\caption{$T$ dependence of 
(a) the resistance $R$,
(b) the sensitivity $S \equiv |dR/dT|$,
(c) the temperature coefficient of resistance (TCR) $S_{\rm T} \equiv |(1/R)dR/dT|$, and 
(d) the dimensionless sensitivity $S_{\rm D} \equiv |(T/R)dR/dT| = |d({\rm ln} R)/d({\rm ln} T)|$ for the RuO$_2$$-$AlO$_x$ films with different RuO$_2$ content fabricated on thermally-oxidized Si substrates. The films are patterned into a Hall-bar shape having the length, width, and thickness of $1.0~\textrm{mm}$, $0.5~\textrm{mm}$, and $\sim 100~\textrm{nm}$, respectively, by co-sputtering RuO$_2$$-$AlO$_x$ through a metal mask. 
In (c) and (d), the $S_{\rm T}\left(T \right)$ and $S_{\rm D}\left(T \right)$ results for the RuO$_2$$-$AlO$_x$ thermometer film on the Pt/YIG sample are coplotted (red star marks).
}
\label{fig:3}
\end{figure} 
%%%%%%%%%%%%%%%%%%%%%%%%%%%%%%%%%%%%%%%%%%%%%%%%%%%%%%%%%%%%%%%%%%%
%%%%%%%%%%%%%%%%%%%%%%%%%%%%%%%%%%%%%%%%%%%%%%%%%%%%%%%%%%%%%%%%%% 
%
%
%
%
%
We here discuss the $T$-dependent thermometer characteristics of the RuO$_2$$-$AlO$_x$ films.  
Figure \ref{fig:3} shows the $T$ dependence of   
(a) the resistance $R$, (b) the sensitivity $S \equiv |dR/dT|$,
(c) the temperature coefficient of resistance (TCR) $S_{\rm T} \equiv |(1/R)dR/dT|$, and (d) the dimensionless sensitivity $S_{\rm D} \equiv |(T/R)dR/dT| = |d({\rm ln} R)/d({\rm ln} T)|$ for the RuO$_2$$-$AlO$_x$ films. 
Here, the sensitivity $S$ is an essential quantity when the thermometer is used as an actual temperature-sensor device in its original form. 
The TCR $S_{\rm T}$ is the normalized sensitivity $S$ by the measured resistance $R$, given that $S$ is geometry dependent (i.e., $dR/dT$ scales with $R$) \cite{LakeShore}. 
The dimensionless sensitivity $S_{\rm D}$ is a measure often used to compare the performance of the thermometers made of different materials, regardless of their size \cite{Harris2021RevSciInstrum,Scott2022JAP,LakeShore,Courts2003AIPConfProc}.
%
%$\leq 41\%$
For the present RuO$_2$$-$AlO$_x$ films with low RuO$_2$ content ($< 40\%$), the sensitivity $S$ takes a high value on the order of $10^4-10^6~\Omega/\textrm{K}$ below $\sim 10~\textrm{K}$. 
For such a low-$T$ range, however, their resistance $R$ values are highly enhanced, and exceed $1~\textrm{M}\Omega$ at $2~\textrm{K}$, which is too high to use such films as thermometers in their original dimensions below  the liquid-$^4$He temperature.  
Besides, their TCR values start to show a saturation behavior by decreasing $T$ in such a low-$T$ environment. 
By contrast, the RuO$_2$$-$AlO$_x$ film with the RuO$_2$ content of 41\% (fabricated under $P_{{\rm RuO}_2} = 28~\textrm{W}$ and $P_{{\rm AlO}_x} = 25~\textrm{W}$) shows a moderate $R$ ($S$) value of $10^4-10^5 ~\Omega$ ($10^4-10^5~\Omega/\textrm{K}$) and the best TCR characteristic of $\sim 100\% ~\textrm{K}^{-1}$ around $2~\textrm{K}$. 
We therefore adopt its growth condition for our SPE device.  
Overall, the $S$, TCR, and $S_{\rm D}$ values of the present  RuO$_2$$-$AlO$_x$ films are comparable to those of commercially available Cernox\texttrademark ~zirconium oxy-nitride sensors \cite{LakeShore,Courts2003AIPConfProc}, carbon composites \cite{Clement1952RevSciInstrum,Lawless1972RevSciInstrum,Pobell-textbook}, and AuGe films \cite{Scott2022JAP} commonly used at a similar $T$ range.
\subsection{Observation of SPE based on RuO$_2$$-$AlO$_x$ on-chip thermometer} \label{sec:SPE_result}
\begin{figure}[htb]
\begin{center}
\includegraphics[width=8.5cm]{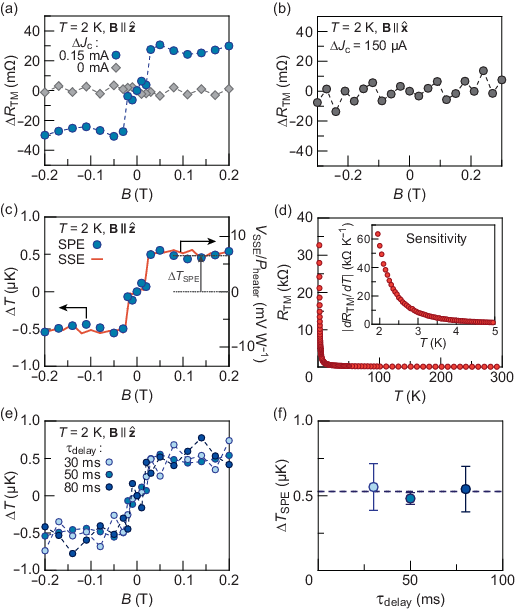}
\end{center}
\caption{
(a) $B$ dependence of the SPE-induced $\Delta R_{\rm TM}$ at $T = 2~\textrm{K}$ and $B \leq 0.2~\textrm{T}$ (${\bf B} \;||\; \mathbf{\hat{z}}$) under $\Delta J_{\rm c}=0.15$ and $0.00~\textrm{mA}$ and $\tau_{\rm delay}=50~\textrm{ms}$. 
The dashed lines connect adjacent plots.
(b) $B$ dependence of $\Delta R_{\rm TM}$ for ${\bf B} \;||\; \mathbf{\hat{x}}$ ($B \leq 0.3~\textrm{T}$) under $\Delta J_{\rm c}=0.15~\textrm{mA}$. Note that the applied $B$ is larger than the out-of-plane (${\bf B} \;||\; \mathbf{\hat{x}}$) saturation field for bulk YIG, which is $\sim 0.2~\textrm{T}$ \cite{Kikkawa2017PRB}.
(c) Comparison between the $B$ dependence of the SPE-induced  temperature change $\Delta T$ (blue filled circles) and the SSE-induced voltage normalized by heating power $V_{\rm SSE}/P_{\rm heater}$ (orange solid curve) at $T=2~\textrm{K}$ and ${\bf B} \;||\; \mathbf{\hat{z}}$. 
The SPE data shown here is the same as that plotted in (a), but the left vertical axis is converted from $\Delta R_{\rm TM}$ to $\Delta T$ via the $R_{\rm TM}$--$T$ calibration curve plotted in (d). 
For details of the SSE measurement, see Sec. \ref{sec:Method_SPE-SSE}.
%For the SSE measurement, an a.c. charge current with the intensity of $I_{\rm rms} = 5.48 ~\mu{\rm A}$ and the frequency of $\omega/2\pi = 13.423~\textrm{Hz}$ was applied to the RuO$_2$-AlO$_x$ thermometer film, and the resultant second harmonic voltage in the Pt layer was detected by a lock-in amplifier. 
(d) $T$ dependence of $R_{\rm TM}$ (main) and $|dR_{\rm TM}/dT|$ (inset) for the RuO$_2$$-$AlO$_x$ film on the Pt/YIG sample.
(e) $B$ dependence of the SPE-induced $\Delta T$ at $T = 2~\textrm{K}$ under $\Delta J_{\rm c}=0.15~\textrm{mA}$ and several $\tau_{\rm delay}$ values. 
(f) $\tau_{\rm delay}$ dependence of the magnitude of the SPE-induced temperature change $\Delta T_{\rm SPE}$, where $\Delta T_{\rm SPE}$ is evaluated by averaging the $\Delta T$ values for $0.08~\textrm{T} \leq B \leq 0.2~\textrm{T}$ [see also (c)]. The dashed line represents the averaged value.
All the $\Delta R_{\rm TM}$ and $\Delta T$ data were anti-symmetrized with respect to the magnetic field $B$. 
}
\label{fig:4}
\end{figure}

We are now in a position to demonstrate a cryogenic SPE in the Pt/YIG sample based on the RuO$_2$$-$AlO$_x$ on-chip thermometer. 
Figure \ref{fig:4}(a) shows the $B$ dependence of the RuO$_2$$-$AlO$_x$ resistance change $\Delta R_{\rm TM}$ measured at $T = 2~\textrm{K}$ and a low-$B$ range of $\left\vert B\right\vert \leq 0.2~\textrm{T}$. 
With the application of the charge current $\Delta J_{\rm c} (=0.15~\textrm{mA})$ to the Pt film, a clear $\Delta R_{\rm TM}$ signal appears with a magnitude saturated at $\sim 30~\textrm{m}\Omega$\cite{Comment_undulation} and its sign changes depending on the ${\bf B}~(||\pm\mathbf{\hat{z}})$ direction.  
The signal disappears either when $\Delta J_{\rm c}$ is essentially zero [gray diamonds in Fig. \ref{fig:4}(a)] or when ${\bf B}$ is applied perpendicular to the Pt/YIG interface (${\bf B} \;||\pm\mathbf{\hat{x}}$) [Fig. \ref{fig:4}(b)]. 
We also confirmed that the $B$ dependence of $\Delta R_{\rm TM}$ is consistent with that of the SSE in the identical Pt/YIG device [see Fig. \ref{fig:4}(c)]. 
These are the representative features of the SPE
\cite{Daimon2016NatCommun,Daimon2017PRB,Itoh2017PRB,Yagmur2018JPhysD,Sola2019SciRep,Yahiro2020PRB,Yamazaki2020PRB,Daimon2020APEX,Uchida2021Review,Uchida2021JPSJ}.  
Furthermore, the sign of $\Delta R_{\rm TM}$ agrees with the SPE-induced temperature change\cite{Daimon2016NatCommun,Daimon2017PRB}. 
As shown in Fig. \ref{fig:4}(a), the measured $\Delta R_{\rm TM}$ value is positive for $B>0$, 
meaning that the resistance $R_{\rm TM}$ increases (decreases) when ${\bf J}_{\rm c} \;||+\mathbf{\hat{y}}$ (${\bf J}_{\rm c} \;||-\mathbf{\hat{y}}$), for which the orientation of the SHE-induced magnetic moment at the interfacial Pt layer is  
$\delta {\bf m}_{\rm s} \;||-\mathbf{\hat{z}}$ ($\delta {\bf m}_{\rm s} \;||+\mathbf{\hat{z}}$) in Fig. \ref{fig:1}(a). 
According to the negative TCR of the RuO$_2$$-$AlO$_x$ film, this implies that the temperature of the Pt film decreases (increases) when $\delta {\bf m}_{\rm s} \;||-\mathbf{\hat{z}}$ ($\delta {\bf m}_{\rm s} \;||+\mathbf{\hat{z}}$) under ${\bf M} \;||\; {\bf B} \;||+\mathbf{\hat{z}}$.  
This correspondence between the sign of the temperature change $\Delta T$ and the relative orientation of $\delta {\bf m}_{\rm s}$ with respect to ${\bf M}$ is consistent with the scenario of the SPE described in Sec. \ref{sec:Method_SPE-SSE}. We thus conclude that we succeeded in measuring a cryogenic SPE using the RuO$_2$$-$AlO$_x$ on-chip thermometer film. \par
%
%
%%%%%%%%%%%%%%%%%%%%%%%%%%%%%%%%%%%%%%%%%%%%%%%%%%%%%%%%%%%%%%%%%%%
%%%%%%%%%%%%%%%%%%%%%%%%%%%%%%%%%%%%%%%%%%%%%%%%%%%%%%%%%%%%%%%%%% 
% \left(B \right)
%
% 
%$325~\textrm{K} \geq T \geq 250~\textrm{K}$ ($225~\textrm{K} \geq T \geq 100~\textrm{K}$)
%
%
%
%
%
%
%
%The inset shows the $B$ dependence of $R_{\rm TM}$ at $T=2~\textrm{K}$ and $B \leq 10~\textrm{T}$ (${\bf B} \;||\; \mathbf{\hat{z}}$). 
%
%
%
%
%
To convert the $\Delta R_{\rm TM}$ value to the temperature change $\Delta T$, we measured the $R_{\rm TM}$--$T$ curve for the RuO$_2$$-$AlO$_x$ film. As shown in  Fig. \ref{fig:4}(d), similar to the results described in Sec. \ref{sec:Electric_conduction}, its resistance $R_{\rm TM}$ increases dramatically with decreasing $T$ at low temperatures,  and the sensitivity $S=|dR_{\rm TM}/dT|$ is as large as $55.3~\textrm{k}\Omega/\textrm{K}$ at 2 K [The TCR $S_{\rm T}$ and dimensionless sensitivity $S_{\rm D}$ for the film are plotted in Figs. \ref{fig:3}(c) and \ref{fig:3}(d), respectively, and $\rho$ and $|d\rho/dT|$ are plotted in Figs. \ref{fig:rhoT_slope}(a) and \ref{fig:rhoT_slope}(b) in APPENDIX B, respectively, together with the results for the RuO$_2$$-$AlO$_x$ films grown on thermally-oxidized Si substrates].
In Fig. \ref{fig:4}(c), we replot the $B$ dependence of the SPE in units of the temperature change $\Delta T ~(= \Delta R_{\rm TM}/S)$ using the above $S$ value. 
We evaluate the magnitude of the SPE-induced temperature change to be $\Delta T_{\rm SPE} = 482 \; \pm \; 39~\textrm{nK}$, by averaging the $\Delta T$ values for $0.08~\textrm{T} \leq B \leq 0.2~\textrm{T}$, at which the magnetization ${\bf M}$ of the YIG slab fully orients along the ${\bf B}$ direction\cite{Uchida2015PRB} [see the dashed line in Fig. \ref{fig:4}(c)]. 
The standard deviation of $39~\textrm{nK}$ shows that our measurement scheme based on the RuO$_2$$-$AlO$_x$ on-chip thermometer can resolve an extremely small $\Delta T$ on the order of several tens of nK (which is a value achieved by repeating the process of the $J_{\rm c}$-polarity change of $7 \times 10^4$ times at each $B$). The $\Delta T$ resolution is much higher than that reported in the previous SPE measurements based on lock-in thermography, lock-in thermoreflectance, and thermocouples, for which the typical resolution is $100$, $10-100$, and $5~\mu\textrm{K}$, respectively\cite{Yahiro2020PRB,Uchida2021JPSJ,Takahagi2023APL}.  
We found that the magnitude of $\Delta T_{\rm SPE}$ normalized by the charge-current density $\Delta j_{\rm c}$ applied to the Pt wire is $\Delta T_{\rm SPE}/\Delta j_{\rm c} = 3.2 \times 10^{-15}~\textrm{Km}^2/\textrm{A}$, which is two orders of magnitude smaller than the corresponding value for Pt/YIG systems measured at room temperature \cite{Daimon2017PRB,Itoh2017PRB}. 
The low-$T$ signal reduction of the SPE is consistent with that found in the SSE \cite{Kikkawa2015PRB,Jin2015PRB,Kikkawa2016JPSJ,Oyanagi2020AIPAdv}, and is attributed mainly to the reduction of the thermally activated magnons contributing to these phenomena at cryogenic temperatures. 
Besides, there can be a finite temperature gradient across the insulating AlO$_x$ film, between the Pt and RuO$_2$$-$AlO$_x$ layers, resulting in further decrease of the detected $\Delta T$ signal.
We also measured the delay time $\tau_{\rm delay}$ dependence of the SPE and found that the $\Delta T_{\rm SPE}$ takes almost the same value in the present $\tau_{\rm delay}$ range ($30~\textrm{ms} \leq \tau_{\rm delay} \leq 80~\textrm{ms}$) [see Figs. \ref{fig:4}(e) and \ref{fig:4}(f)], showing that all the data were obtained under the steady-state condition \cite{Itoh2017PRB,Yamazaki2020PRB}. \par
\begin{figure}[htb]
\begin{center}
\includegraphics[width=8.5cm]{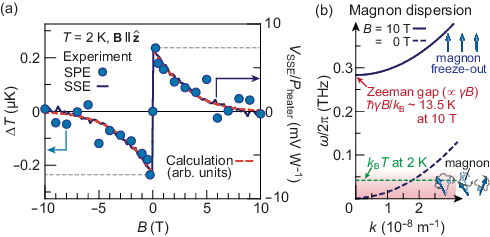}
\end{center}
\caption{
(a) Comparison between the high magnetic field $B$ response of the SPE-induced temperature change $\Delta T$ (blue filled circles) and the SSE-induced voltage normalized by heating power $V_{\rm SSE}/P_{\rm heater}$ (blue solid curve) at $T=2~\textrm{K}$ and $B \leq 10~\textrm{T}$ (${\bf B} \;||\; \mathbf{\hat{z}}$). 
The SPE data was obtained under $\Delta J_{\rm c}=0.15~\textrm{mA}$ and $\tau_{\rm delay} = 50~\textrm{ms}$. 
For details of the SSE measurement, see Sec. \ref{sec:Method_SPE-SSE}. The orange dashed curve shows the numerically calculated result based on Eq. (\ref{equ:B-dep-Calc}) for $T=2~\textrm{K}$.
(b) Magnon dispersion relations for YIG\cite{Kikkawa2016PRL} at $B = 0$ and $10~\textrm{T}$, at which the magnon-excitation gap values are $\sim 0$ and $13.5~\textrm{K}$ in units of temperature, respectively, where $k$ represents the wavenumber. The thermal energy ($k_{\rm B}T$) level of $2~\textrm{K}$ is also plotted with a green dashed line, above which thermal excitation is exponentially suppressed. 
%At $B = 0~\textrm{T}$, magnons can be thermally excited, whereas at $B = 14~\textrm{T}$ magnons are frozen out because $\hbar \omega \gg k_{\rm B}T$.
}
\label{fig:5}
\end{figure}

We also explored the high magnetic field response of the SPE signal. 
Figure \ref{fig:5}(a) displays the $\Delta T$ versus $B$ data measured at $T=2~\textrm{K}$ and $B \leq 10~\textrm{T}$ (${\bf B} \;||\; \mathbf{\hat{z}}$). We found that $\Delta T$ exhibits a maximum at a low $B$ ($\lesssim 0.2~\textrm{T}$) and, by increasing $B$, gradually decreases and is eventually suppressed. The $B$ dependence of the SPE agrees well with that of the SSE measured with the identical device [see Fig. \ref{fig:5}(a)].
We note that the magnetoresistance (MR) ratio of the RuO$_2$$-$AlO$_x$ film is as small as $\sim 3.7 \%$ for $B \leq 10~\textrm{T}$ at $T = 2~\textrm{K}$, so that the device can be used reliably under the high-$B$ range.  
The observed $\Delta T \left(B \right)$ feature is explained in terms of the suppression of magnon excitations by the Zeeman effect, as established in the previous SSE research \cite{Kikkawa2015PRB,Jin2015PRB,Kikkawa2016JPSJ,Oyanagi2020AIPAdv,Kikkawa2021NatCommun} [see Fig. \ref{fig:5}(b)]. 
By increasing $B$, the magnon dispersion shifts toward high frequencies due to the Zeeman interaction ($\propto \gamma B$). 
At $B = 10~\textrm{T}$, the Zeeman energy $\hbar \gamma B$ is $\sim 13.5~\textrm{K}$ in units of temperature, which is greater than the thermal energy $k_{\rm B}T$ at 2 K [see Fig. \ref{fig:5}(b)], resulting in an insignificant value of the Boltzmann factor: ${\rm exp}(-\hbar \gamma B/k_{\rm B}T) \sim 10^{-3} \ll 1$, where $\gamma$ and $\hbar$ represent the gyromagnetic ratio and Dirac constant, respectively. 
Therefore, the thermal magnons that can contribute to the SPE at a low $B$ are gradually suppressed with the increase of $B$ and, at $B \sim 10~\textrm{T}$, are hardly excited by the strong Zeeman gap in the magnon spectrum [Fig. \ref{fig:5}(b)], which leads to the suppression of the SPE in the low-$T$ and high-$B$ environment. 
%%%%%%%%%%%%%%%%%%%%%%%%%%%%%%%%%%%%%%%%%%%%%%%%%%%%%%%%%%%%%%%%%%
%%%%%%%%%%%%%%%%%%%%%%%%%%%%%%%%%%%%%%%%%%%%%%%%%%%%%%%%%%%%%%%%%% 
%
%$\Delta T \left(B \right)$
We also compared the experimental result with a calculation for the interfacial heat current induced by the SPE $J_{\rm q}^{\rm SPE}$ and spin current induced by the SSE $J_{\rm s}^{\rm SSE}$, which are expressed as 
\begin{equation}
\begin{split}\label{equ:B-dep-Calc}
J_{\rm q}^{\rm SPE} & \propto  \int \frac{d^3k}{(2\pi)^3}  \: \omega^2 \frac{\partial n_{\rm BE}}{\partial \omega}, \\
J_{\rm s}^{\rm SSE} & \propto  - \int \frac{d^3k}{(2\pi)^3}  \: \omega T \frac{\partial n_{\rm BE}}{\partial T}, 
\end{split}
\end{equation}
respectively \cite{Kikkawa2015PRB,Adachi2013review,Bender2015PRB,Cornelissen2016PRB,Ohnuma2017PRB}. Here, $\omega = D_{\rm ex}k^2 + \gamma B$ is the parabolic magnon dispersion for YIG with the stiffness constant of $D_{\rm ex} = 7.7 \times 10^{-6}~\textrm{m}^2/\textrm{s}$ \cite{Kikkawa2016PRL} and $n_{\rm BE}=[{\rm exp}(\hbar \omega/k_{\rm B}T)-1]^{-1}$ is the Bose$-$Einstein distribution function. Note that the relation $\omega \partial n_{\rm BE}/\partial \omega = -T \partial n_{\rm BE}/\partial T$ ensures the Onsager reciprocity between the SSE and SPE\cite{Ohnuma2017PRB}, which makes the above expressions to be of the same form in terms of the $B$ dependence. %\cite{Itoh2017PRB,Yahiro2020PRB}. 
As shown by the orange dashed curve in Fig. \ref{fig:5}(a), the calculated result based on Eq. (\ref{equ:B-dep-Calc}) well reproduces the experiment.  
This result further supports the origin of the measured $\Delta T$ signal and provides additional clues for further understanding of the physics of the SPE. \par
% 
%
%%%%%%%%%%%%%%%%%%%%%%%%%%%%%%%%%%%%%%%%%%%%%%%%%%%%
\section{CONCLUSIONS}
%\section{IV.~~CONCLUSIONS}
%%%%%%%%%%%%%%%%%%%%%%%%%%%%%%%%%%%%%%%%%%%%%%%%%%%%
% 
In this study, we have fabricated RuO$_2$$-$AlO$_x$ films by means of a d.c. co-sputtering technique and characterized their electrical conduction and sensitivity at low temperatures.  
The sensitivity was found to be tuned simply by the relative sputtering power applied for the RuO$_2$ and Al targets, and the TCR value reaches $\sim 100\%~\textrm{K}^{-1}$ for the RuO$_2$$-$AlO$_x$ films with the moderate RuO$_2$ content ($\gtrsim 41\%$). 
\begin{figure}[htb]
\begin{center}
\includegraphics[width=5.5cm]{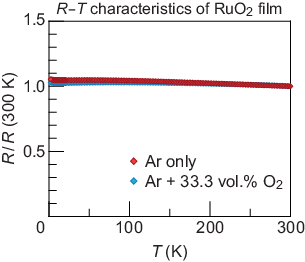}
\end{center}
\caption{ 
$T$ dependence of $R/R(T=300~\textrm{K})$ for the pure RuO$_2$ films grown under only Ar gas flow and also under a large amount of O$_2$ gas flow (Ar + 33.3 vol.\% O$_2$), for which the resistivity $\rho$ values at $T = 300~\textrm{K}$ are evaluated as $3.21 \times 10^{-4}$ and $3.96 \times 10^{-4} ~\Omega \textrm{cm}$, respectively.  
}
\label{fig:RT_RuO2}
\end{figure}
By using the RuO$_2$$-$AlO$_x$ film as an on-chip micro-thermometer, we successfully measured the SPE-induced temperature change $\Delta T$ in a Pt-film/YIG-slab system at the low temperature of 2 K based on the so-called Delta method, which can resolve an extremely small $\Delta T$ value of several tens of nK. 
We also measured the high $B$ response of the SPE at $T = 2~\textrm{K}$ up to $B = 10~\textrm{T}$, and found that, by increasing $B$, the SPE signal gradually decreases and is eventually suppressed. 
The $B$ dependence can be interpreted in terms of the field-induced freeze-out of magnons due to the Zeeman-gap opening in the magnon spectrum of YIG. 
We anticipate that our experimental methods based on an on-chip  thin-film thermometer will be useful for exploring low-$T$ thermoelectric heating/cooling effects in various types of micro devices, including a system based on two-dimensional van der Waals materials \cite{Kanahashi2019review,Li2020review,Wang2022ASCNano_review}. 
Besides, with an appropriate optimization of the resistance and sensitivity of the RuO$_2$$-$AlO$_x$ films by controlling the content of RuO$_2$, our results can be extended toward even lower temperature ranges below 1 K, where they can be used to detect unexplored cryogenic spin caloritronic effects driven by nuclear and quantum spins.  
\par 
%
%%%%%%%%%%%%%%%%%%%%%%%%%%%%%%%%%%%%%%%%%%%%%%%%%%%%
\section*{ACKNOWLEDGMENTS}
%%%%%%%%%%%%%%%%%%%%%%%%%%%%%%%%%%%%%%%%%%%%%%%%%%%%
%
%
We thank S. Daimon, R. Yahiro, J. Numata, K. K. Meng, H. Arisawa, T. Makiuchi, and T. Hioki for valuable discussions. 
This work was supported by 
JST-CREST (JPMJCR20C1 and JPMJCR20T2), 
Grant-in-Aid for Scientific Research (JP19H05600, JP20H02599, and JP22K18686) and 
Grant-in-Aid for Transformative Research Areas (JP22H05114) from JSPS KAKENHI, 
MEXT Initiative to Establish Next-generation Novel Integrated Circuits Centers (X-NICS) (JPJ011438), Japan, 
Murata Science Foundation, 
Daikin Industries, Ltd, 
and Institute for AI and Beyond of the University of Tokyo. \par
%
%
%
%
%%%%%%%%%%%%%%%%%%%%%%%%%%%%%%%%%%%%%%%%%%%%%%%%%%%%%%%%%%%%%%%%%%%%%% 
\section*{APPENDIX A: ELECTRICAL CONDUCTION IN PURE RUO$_2$ FILMS} \label{sec:Appendix-RuO2-RvsT}
\begin{figure}[htb]
\begin{center}
\includegraphics[width=8.5cm]{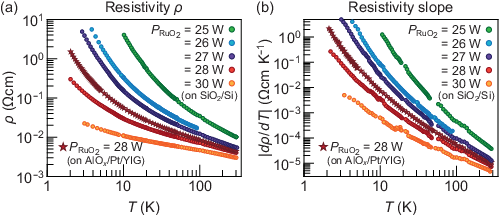}
\end{center}
\caption{ 
$T$ dependence of 
(a) the resistivity $\rho$ and
(b) its slope $|d\rho/dT|$ 
for the RuO$_2$$-$AlO$_x$ films on thermally-oxidized Si substrates  (filled circles) and  on the Pt/YIG sample (red star marks) grown under the several d.c. sputtering power for the RuO$_2$ target ($P_{{\rm RuO}_2}$) and the fixed d.c. power for the Al target ($P_{{\rm AlO}_x} = 25~\textrm{W}$). 
Note that $S_{\rm T} \equiv |(1/R)dR/dT| = |(1/\rho)d\rho/dT|$ and $S_{\rm D} \equiv |(T/R)dR/dT| = |(T/\rho)d\rho/dT|$, the $T$ dependences of which are shown in Figs. \ref{fig:3}(c) and \ref{fig:3}(d), respectively.  
}
\label{fig:rhoT_slope}
\end{figure} 
To check the effect of O$_2$ gas introduction during sputtering on the RuO$_2$ film, we also fabricated pristine (polycrystalline) RuO$_2$ films under only Ar gas flow and also under a large amount of O$_2$ gas flow (Ar + 33.3 vol.\% O$_2$, 
which is $\sim$ 5 times greater than that used for the RuO$_2$$-$AlO$_x$ deposition) and measured their $R$--$T$ curves. 
Here, the RuO$_2$ films were patterned into a Hall-bar shape having the length, width, and thickness of $2.0~\textrm{mm}$, $0.3~\textrm{mm}$, and $\sim 10~\textrm{nm}$, respectively, by sputtering RuO$_2$ through a metal mask. Figure \ref{fig:RT_RuO2} shows the $T$ dependence of $R$ normalized by the value at 300 K for each RuO$_2$ film. For both the films, the $R$--$T$ curve shows almost the same characteristics; $R$ gradually increases with decreasing $T$ and the $R(T)/R(300~\textrm{K})$ value at $T = 2 ~\textrm{K}$ (the temperature of interest) deviates only $\sim$ 2\% with each other. This result shows that the effect of oxygen on the RuO$_2$ deposition does not play an essential role in the $R$ versus $T$ characteristics of RuO$_2$. \par 
%
%
%%%%%%%%%%%%%%%%%%%%%%%%%%%%%%%%%%%%%%%%%%%%%%%%%%%%%%%%%%%%%%%%%%%%%%% 
\section*{APPENDIX B: COMPARISON OF $\rho$$-$$T$ CURVES BETWEEN RUO$_2$$-$ALO$_x$ FILMS ON SIO$_2$/SI SUBSTRATES AND ON PT/YIG DEVICE} \label{sec:Appendix-RuO2-rhovsT}
Figures \ref{fig:rhoT_slope}(a) and \ref{fig:rhoT_slope}(b) show the double logarithmic plot of (a) the resistivity $\rho$ and (b) its slope $|d\rho/dT|$ versus temperature $T$ for the RuO$_2$$-$AlO$_x$ films on thermally-oxidized Si substrates (filled circles) and on the Pt/YIG sample (red star marks) grown under the several d.c. sputtering power for the RuO$_2$ target ($P_{{\rm RuO}_2}$) and the fixed d.c. power for the Al target ($P_{{\rm AlO}_x} = 25~\textrm{W}$). 
Although a small deviation of the $\rho$ and $|d\rho/dT|$ values is observed even under the same growth condition depending on the substrate layer (i.e., SiO$_2$/Si or Pt/YIG), the overall $T$ dependent feature agrees well with each other. Note that the substrate-dependent difference in the $\rho$ and $|d\rho/dT|$ values does not have a significant impact on the observation of the cryogenic SPE, if the sensitivity is large enough for its detection. \par
%between the RuO$_2$$-$AlO$_x$ films grown on thermally-oxidized Si  substrates and on the Pt/YIG device.
%
%
%
%
%
%
%

%\newpage
%
%
%
%
%
%
%
%
%%%%%%%%%%%%%%%%%%%%%%%%%%%%%%%%%%%%%%%%%%%%%%%%%%%%%%%%%%%%%%%%%%
%%%%%%%%%%%%%%%%%%%%%%%%%%%%%%%%%%%%%%%%%%%%%%%%%%%%%%%%%%%%%%%%%% 
%

\begin{thebibliography}{99}
\bibitem{Comment_maintenance} Except for this procedure, we do not need to be extremely careful about the maintenance of the RuO$_2$ and Al targets. As a side note, during the pre-sputtering process, RuO$_2$ thin flakes that are easily peeled off may form on the shutter in the sputtering chamber, so we need to clean it regularly to prevent these flakes from falling on the RuO$_2$ target.
\bibitem{Comment_different-T} Since the SPE measurement in this study was intended to be performed at 2 K, the conditions for the thermometer fabrication were chosen to realize high sensitivity at 2 K. Our method can be extended to different temperatures by making thermometer films with high sensitivity at the targeted temperature range. However, as the temperature is lowered toward zero kelvin, the SPE-induced temperature difference is expected to decrease because the population of thermal magnons in YIG is suppressed.
\bibitem{Comment_Joule-heating} The Joule-heating-induced temperature increase under a typical $\Delta J_{\rm c}$ value of 0.15 mA is less than 10 mK, which was confirmed by measuring the $R_{\rm TM}$ value of the RuO$_2$$-$AlO$_x$ film under the d.c. current of 0.15 mA applied to the Pt wire at $T=2~\textrm{K}$ and by comparing it with the $R_{\rm TM}$--$T$ calibration curve shown in Fig. \ref{fig:4}(d). The current value of 0.15 mA was adopted such that the sample's temperature increase is less than 1\% compared to the base temperature of 2 K and also that the SPE signal can be measured with a sufficient signal-to-noise ratio.
\bibitem{Comment_frequency} For the delay time of $\tau_{\rm delay} = 50~\textrm{ms}$ and the data acquisition time $\tau_{\rm sens} = 20~\textrm{ms}$, the frequency of the applied square-wave current is estimated to be $f = 1/[2(\tau_{\rm delay} + \tau_{\rm sens})] \sim 7~\textrm{Hz}$.
\bibitem{Comment_undulation} A typical error at each field is evaluated to be $9.8~\textrm{m}\Omega$ for the data under $\Delta J_{\rm c}=0.15~\textrm{mA}$, so the slight undulation behavior of $\Delta R_{\rm TM}$ above $B=0.05~\textrm{T}$ is attributed to the uncertainty of the measurement.
\end{thebibliography}
\end{document}